\input psfig.tex
\documentstyle{mn}

\newif\ifAMStwofonts

\ifoldfss
  \ifCUPmtlplainloaded \else
    \NewTextAlphabet{textbfit} {cmbxti10} {}
    \NewTextAlphabet{textbfss} {cmssbx10} {}
    \NewMathAlphabet{mathbfit} {cmbxti10} {} 
    \NewMathAlphabet{mathbfss} {cmssbx10} {} 
  \fi
  \ifAMStwofonts
    \ifCUPmtlplainloaded \else
      \NewSymbolFont{upmath} {eurm10}
      \NewSymbolFont{AMSa} {msam10}
      \NewMathSymbol{\upi}     {0}{upmath}{19}
      \NewMathSymbol{\umu}     {0}{upmath}{16}
      \NewMathSymbol{\upartial}{0}{upmath}{40}
      \NewMathSymbol{\leqslant}{3}{AMSa}{36}
      \NewMathSymbol{\geqslant}{3}{AMSa}{3E}

    \fi
  \fi
\fi 

\ifnfssone
  \newmathalphabet{\mathit}
  \addtoversion{normal}{\mathit}{cmr}{m}{it}
  \addtoversion{bold}{\mathit}{cmr}{bx}{it}
  \newmathalphabet{\mathbfit} 
  \addtoversion{normal}{\mathbfit}{cmr}{bx}{it}
  \addtoversion{bold}{\mathbfit}{cmr}{bx}{it}
  \newmathalphabet{\mathbfss} 
  \addtoversion{normal}{\mathbfss}{cmss}{bx}{n}
  \addtoversion{bold}{\mathbfss}{cmss}{bx}{n}
  \ifAMStwofonts
    \ifCUPmtlplainloaded \else
      \UseAMStwoboldmath
      \makeatletter
      \new@mathgroup\upmath@group
      \define@mathgroup\mv@normal\upmath@group{eur}{m}{n}
      \define@mathgroup\mv@bold\upmath@group{eur}{b}{n}
      \edef\UPM{\hexnumber\upmath@group}
      \new@mathgroup\amsa@group
      \define@mathgroup\mv@normal\amsa@group{msa}{m}{n}
      \define@mathgroup\mv@bold\amsa@group{msa}{m}{n}
      \edef\AMSa{\hexnumber\amsa@group}
      \makeatother
      \mathchardef\upi="0\UPM19
      \mathchardef\umu="0\UPM16
      \mathchardef\upartial="0\UPM40
      \mathchardef\leqslant="3\AMSa36
      \mathchardef\geqslant="3\AMSa3E
    \fi
  \fi
\fi 

\ifnfsstwo
  \DeclareMathAlphabet{\mathbfit}{OT1}{cmr}{bx}{it}
  \SetMathAlphabet\mathbfit{bold}{OT1}{cmr}{bx}{it}
  \DeclareMathAlphabet{\mathbfss}{OT1}{cmss}{bx}{n}
  \SetMathAlphabet\mathbfss{bold}{OT1}{cmss}{bx}{n}
  \ifAMStwofonts
    \ifCUPmtlplainloaded \else
      \DeclareSymbolFont{UPM}{U}{eur}{m}{n}
      \SetSymbolFont{UPM}{bold}{U}{eur}{b}{n}
      \DeclareSymbolFont{AMSa}{U}{msa}{m}{n}
      \DeclareMathSymbol{\upi}{0}{UPM}{"19}
      \DeclareMathSymbol{\umu}{0}{UPM}{"16}
      \DeclareMathSymbol{\upartial}{0}{UPM}{"40}
      \DeclareMathSymbol{\leqslant}{3}{AMSa}{"36}
      \DeclareMathSymbol{\geqslant}{3}{AMSa}{"3E}
    \fi
  \fi
\fi 

\ifCUPmtlplainloaded \else
  \ifAMStwofonts \else 
    \def\upi{\pi}
    \def\umu{\mu}
    \def\upartial{\partial}
  \fi
\fi

\title{A fireworks model for Gamma-Ray Bursts}

\author[G.Barbiellini, A.Celotti and F.Longo] { Guido
Barbiellini$^{1}$, Annalisa Celotti$^{2}$ and Francesco Longo$^{1}$ \\
(1) Department of Physics and INFN, via Valerio 2, I-34100 Trieste,
Italy \\ (2) SISSA, via Beirut 2-4, I-34014 Trieste, Italy} 

\pagerange{\pageref{firstpage}--\pageref{lastpage}} 

\begin{document} 

\maketitle

\begin{abstract}
  The energetics of the long duration GRB phenomenon is compared with
  models of a rotating Black Hole (BH) in a strong magnetic field
  generated by an accreting torus. A rough estimate of the energy
  extracted from a rotating BH with the Blandford-Znajek mechanism is
  obtained with a very simple assumption: an inelastic collision
  between the rotating BH and the torus. The GRB energy emission is
  attributed to an high magnetic field that breaks down the vacuum
  around the BH and gives origin to a e$^\pm$ fireball. Its subsequent
  evolution is hypothesised, in analogy with the in-flight decay of an
  elementary particle, to evolve in two distinct phases. The first one
  occurs close to the engine and is responsible of energising and
  collimating the shells. The second one consists of a radiation
  dominated expansion, which correspondingly accelerates the
  relativistic photon--particle fluid and ends at the transparency
  time.  This mechanism simply predicts that the observed Lorentz
  factor is determined by the product of the Lorentz factor of the
  shell close to the engine and the Lorentz factor derived by the
  expansion. An anisotropy in the fireball propagation is thus
  naturally produced, whose degree depends on the bulk Lorentz factor
  at the end of the collimation phase.

\end{abstract}


\section{Introduction}

At cosmological distances the observed GRB fluxes imply energies of
order of up to a solar rest-mass ($\sim 10^{54}$ erg), and as they
vary on timescales of the order of milli seconds from causality
arguments these must arise in regions whose size is of the order of
kilometres. This implies that an $e^\pm,\gamma$ fireball must form,
which would expand relativistically.  The fireball is energised and
possibly collimated, mechanically or magnetically, close to the engine
(for reviews see e.g. Piran 1999; Meszaros 2002). Subsequently it
adiabatically expands and accelerates, until the Thomson transparency
is reached (the opacity being determined by either electron--positron
pairs or electrons if the fireball is baryon loaded).  The GRB
phenomenology -- in particular the fast variability and the detection
of $\gamma$--ray emission from an apparently compact region opaque to
electron--positron production via photon--photon interaction -- gives
compelling reasons for the bulk motion of the emitting plasma to be
highly relativistic with Lorentz factors of the order $\Gamma\sim
10^2-10^3$.

The degree of isotropy/collimation of the ejected fireball is however
still unclear. In fact, as the observer only detects $\gamma$--ray
flux from an angle $\sim \Gamma^{-1}$, it is not possible to simply
discriminate between an isotropic and a jet-like structure from the
observed GRB event.  Nevertheless this is in principle possible by
adequate sampling and determination of the behaviour of the light
curves during the afterglow phase: following the deceleration/sideway
expansion of the fireball more and more of the emitting plasma can be
seen and a break (and steepening) in the light curve would appear when
the whole of the volume becomes observable.

Indeed, recently a few GRB afterglows were observed at many
wavelengths and suggest an axisymmetric jet-like structure for the
fireball, thus strongly reducing the estimate of the energetics with
respect to the isotropic case (Frail et al. 2001), although clearly
increasing the required GRB rate.  The temporal decays of the emission
at different frequencies, interpreted according to the fireball model,
suggest jet beaming with opening angles $\theta \sim 3^{\circ}$ (Frail
et al. 2001).  An important inference from these observations is also
that the GRB have a typical energy with little intrinsic spread (Frail
et al. 2001), although alternative possibilities, such as anisotropy
of a collimated fireball, can account for the same observed
phenomenology (Zhang \& Meszaros 2002; Rossi, Lazzati \& Rees 2002).
Found observational trends among timing and spectral properties of GRB
as well as numerical results appear also to favour anisotropic
distributions of energy/velocity in the fireball (Lloyd-Ronning \&
Ramirez-Ruiz 2002; Salmonson \& Galama 2002; Zhang, Woosley \&
MacFayden 2002).

A further important discovery made by $Beppo$SAX, ASCA and Chandra
telescopes, is the presence of iron lines in the X-ray spectra of GRBs
(e.g. Amati et al. 2000; Piro et al. 2000; Antonelli et al. 2000).
This provides a powerful tool to understand the nature and the
environment of GRB primary sources (Vietri et al. 2001; Rees \&
M\'esz\'aros 2000): strong iron lines imply a rich environment which
may be an argument in favour of massive-star progenitor models of GRB
(Woosley 1993, Paczynski 1993; Paczynski 1998; Vietri \& Stella 1998).
These findings have been recently accompanied by the claim of the
observation of a complex of soft X--ray lines by XMM-Newton in the
spectrum of GRB 011211 (Reeves et al. 2002, see also Watson et al.
2002). They suggest in particular that the high temperature derived
from the emitting gas could be interpreted as reheating of pre-ejected
material by the GRB itself.

These observations are in favour of the interpretation of GRBs as a
second step of the residual of the primary explosion (e.g. Vietri \&
Stella 1998): the primary explosion leaves over a compact object that
could be a rotating black hole, at the centre of a rarefied atmosphere
of ejecta. In such scenario it is plausible that the energy extraction
from a rotating BH, through the Blandford-Znajek (BZ) mechanism
(Blandford \& Znajek 1977, Lee et al. 2000), where the external
magnetic field can be supplied by a torus circulating around the BH at
a distance of the order of the Schwarzchild radius R$_{\rm s}$.

In this paper we focus on two aspects of the 'standard' scenario for
the GRB event. The first, developed in Section 2, concerns the
extraction of energy from a rotating compact object and its conversion
into a photon-e$^{\pm}$ fireball. Subsequently, in Section 3, we
suggest that the acceleration and collimation could occur in two
phases, the first one consists in energising and collimating the
shells, the second one of a radiation dominated expansion.  This
mechanism predicts that the observed Lorentz factor is determined by
the product of the Lorentz factor of the shell close to the engine and
the Lorentz factor derived by the expansion, thus naturally giving
rise to an anisotropic fireball. Our conclusions are reported in Section 4.

\section{Gamma-Ray Burst Progenitor}

\subsection{Energetics}
 
As mentioned, Blandford and Znajek have proposed an interaction
between a rotating BH and an accretion disk to explain the energetics
of Active Galactic Nuclei. The same mechanism could be a good
candidate for GRB engines as already pointed out (e.g. Paczynski 1998;
Lee et al 2000). In the BZ mechanism the magnetic field of the
accretion disk acts as a break on the BH and the energy output is
mainly due to the loss of rotational energy. The rotational energy for
a maximally rotating BH of mass $M_{\rm bh}$, with the rotation parameter
$\tilde{a} = {Jc}/{M_{\rm bh}^2 G} = 1$, is 0.29 $M_{\rm bh} c^2$. Even with
the optimal efficiency the available extractable energy for the BZ
mechanism is (Lee et al 2000):
 
$$
E_{\rm BZ} = 0.3 \times 10^{54} \left( \frac{M_{\rm bh}}{M_{\odot}} \right)
\mathrm{erg}.
$$
 
In the following considerations it will be assumed that
a dissipative interaction is at work between the BH and the torus
surrounding it, due to an internal torque. If the short interaction is
treated as an inelastic shock it is possible to apply the angular
momentum conservation law

$$ 
I_{\rm bh}\Omega_{\rm bh} +
I_{\rm t}\Omega_{\rm t} = I\Omega, 
$$
where the subscripts `${bh}$' and `${t}$' refer to the black hole
and torus, respectively, and the right hand side quantities are those
of the final BH slowed down by this interaction. In this
approximation, the loss of rotational and gravitational energy (
  considering the torus approximately at the last stable orbit) can
be derived as

$$
\Delta E_{\rm rot} \simeq \frac{1}{2}I_{\rm bh}\Omega_{\rm bh}^2 \left( 1 -
  \frac{I_{\rm bh}}{I}\right) = 2 M_{\rm bh}^3 \Omega_{\rm bh}^2 \left( 1 -
  \frac{M_{\rm bh}^3}{M^3} \right)\sim 
$$

$$
\sim 2 M_{\rm bh}^3 \Omega_{\rm bh}^2 \left(3 \frac{M_{\rm t}}{M_{\rm bh}}\right) \simeq
\frac{3}{2} \frac{a^2}{R_{\rm bh}^2}M_{\rm t}\simeq
$$

$$
\simeq 3 E_{\rm rot,bh}\frac{M_{\rm t}}{M_{\rm bh}} \sim \frac{3}{8} M_{\rm t} c^2
$$

$$
\Delta E_{\rm g} = \frac{G M_{\rm t}M_{\rm bh}}{R_s}-\frac{G M_{\rm t} M_{\rm bh}}{3 R_{\rm s}}
\simeq \frac{1}{3} M_{\rm t}c^2 \, \, .
$$

The total available energy is therefore $\Delta E_{\rm rot} + \Delta E_{\rm g}
\simeq 0.7 M_{\rm t} c^2$, ranging between $10^{53}- 10^{54}$ erg
for $M_{\rm t} = 0.1 - 1 M_{\odot}$.  In the following it will be assumed
that the energy source of the GRB is the gravitational collapse of a
torus of 0.1 $M_{\odot}$ onto a rotating BH of 10 $M_{\odot}$.

\subsection{Vacuum breakdown}
 
A model for the generation of the GRB fireball is the vacuum breakdown
in the volume close to the polar cap of the BH (Heyl 2001).  A similar
process in the proximity of a charged black hole has been considered
by Ruffini and collaborators (e.g. Ruffini 1998).

The accreted material, which releases its gravitational energy, gives
origin to a variable magnetic field: the field required to explain the
high luminosity of GRB generates an electric field that could break
down the vacuum.

In a recent analysis of the field around the BH Heyl (2001) obtained a
value of $B_c \sim 4.5 \times 10^{13} $ G for the vacuum breakdown in
the ergosphere. The corresponding magnetic energy density is $U_{\rm
  B} \simeq 8\times 10^{25}$ erg cm$^{-3}$.  An estimate of the
electric energy density can be obtained by considering the Wald charge
(Wald 1974), $Q_{\rm w} \sim 2 B a M_{\rm bh}$ (in geometrical units)
$\sim 2\times 10^{16}$ C, corresponding to an electric field (at
$R_{s}$) $E \simeq 2\times 10^{15}$ V cm$^{-1}$ and energy density of
order $U_{\rm e}\simeq 2\times 10^{24}$ erg cm$^{-3}$.
  
While the very same existence of the Wald charge has been questioned
(Shatskiy 2001), similar results are obtained by considering the
voltage drop created by the BZ mechanism

$$
\Delta V = 10^{22} \left( \frac{M_{\rm bh}}{M_{\odot}}\right) \left(
  \frac{B}{10^{15} {\rm G}} \right) \mathrm{V},$$
which in the proximity of the BH corresponds to an electric field
 $$ E = \frac{\Delta V}{2 \pi R} = 5\times 10^{15} \left(\frac{B}{10^{15}{\rm G}}\right)
  \frac{\mathrm{V}}{\mathrm{cm}}. 
$$
Therefore, considering the BZ mechanism to be responsible for
energising GRBs, a magnetic field indeed of order $ B \sim 10^{15}
{\mathrm G}$ can account for the electric field required to break the
vacuum.
Nevertheless, in the following we will assume the description of the
field around the BH obtained by Heyl (Heyl 2001).

In the proximity of the BH is thus possible to generate e$^\pm$ pairs
which could give origin to the GRB fireball, provided a sufficiently
clean environment in order to avoid previous electric field discharge.
This condition can be verified if the relevant matter resides in
the BH and in the rotating torus and the residual density close to the
rotational axis is less than 10$^{9}$ cm$^{-3}$ (Shatskiy
\& Kardashev 2002, Goldreich \& Julian 1969).
Considering a typical electromagnetic field configuration around a
Kerr BH (e.g. Punsly 2001), it is possible that initially the $E$
field generated by the rotation of the BH in the magnetic field of the
torus (e.g. Shatskiy 2001) can actually contribute to clear the
environment of electron-proton plasma.

Note that the recent observation of GRB011211 by XMM (Reeves et al.
2002) reported an absorption edge at 1 keV with optical depth $\tau
\sim 1$. If this evidence will be confirmed by other observations and
assuming a homogeneous environment density, we could put a lower limit
on the particle column density ($10^{23}$ cm$^{-2}$) from the source
to the X-ray photosphere, in favour of a relative clean environment at
least on the jet axis direction.

\subsection{The formation of the fireball}

A magnetic field of the order of $B_{c}$ breaks the vacuum in a volume
$V \sim R_{s}^3$ (cf Heyl 2001).

The number of e$^\pm$ pairs would be 
$$
N_{\rm e^{\pm}} = 2\times (2\pi)^3 \frac{V}{\lambda_{\rm e}^3}
\simeq 10^{51},$$
considering for each e$^\pm$ pair a volume of the
order of $(\lambda_e/2\pi)^3$ where $\lambda_e$ is the electron
Compton wavelength, with a corresponding particle density of
4$\times10^{31}$ e$^{\pm}$ cm$^{-3}$.  This density is evaluated for a
single e$^{\pm}$ pair. For a large population it looks more adequate
to adopt a typical white dwarf density, of order $\sim 4\times
10^{29}$ cm$^{-3}$ (Fermi 1966).

The available magnetic energy density for a field of order of $B_{c}$
implies 
that each outgoing particle gets an energy $\epsilon_0 \sim 10^{-4}
\eta_{\rm acc}$ erg, where $\eta_{\rm acc}$ is the acceleration
efficiency. Its relativistic Lorentz factor is then $\gamma_0 =
\epsilon_0/ m_{\rm e} c^2 \sim 10^2 \eta_{\rm acc}$.

After the formation of the plasmoid the particles undergo three
important processes:

1) Particle acceleration in a time scale
$$t_{\rm acc} \sim \frac{10^2 \eta_{\rm acc} m_{\rm e} c^2}{e \cdot E
  \cdot c } \sim  10^{-19}\eta_{\rm acc}\qquad {\rm s}$$
to acquire an energy
$\sim 10^2 \eta_{\rm acc} m_{\rm e} c^2$ in a electric field of the
order of $2\times 10^{15}$ V/cm.

2) Single particle collimation in the direction of the magnetic field
by synchrotron radiation. The particles momentum components normal to
the magnetic field $p_\perp$ are damped in a time scale
$$
t_{\mathrm{coll}} < \frac{\rho}{c \sin \lambda},
$$
where the curvature radius $\rho$ is of the order of $\sim 3 \times
10^6 E(\mathrm{GeV})/ B(\mathrm{G})$ cm and $\lambda$ is the angle
between the particle motion and the magnetic field.  With the presence
of a magnetic field of the order of $4 \times 10^{13}$ G, the
particles radiate all the energy corresponding to $p_\perp$ in a time
scale
$$
t_{\mathrm{coll}} \sim \frac{\eta_{\rm acc}}{\sin \lambda} 10^{-19}\qquad {\rm s}.
$$
The momentum components perpendicular to field line outside the
plasmoid for all the particles are damped and the plasmoids becomes a
stream of particles with velocity parallel to the external field lines
with $\gamma \sim \gamma_0/3$. As a result the plasmoid travels as a
parallel stream with bulk Lorentz factor
$$\Gamma_1 = \gamma \sim 30\, \eta_{\rm acc}.$$

3) Momenta randomisation on a time scale 
$$t_{\mathrm{rand}} \sim l/c \sim 10^{-12} \eta_{\rm acc}^{-2} \qquad {\rm
  s},$$
where $l$ is the mean free path for e$^\pm$ interaction, $ l =
(\sigma n)^{-1}$, using $\sigma = 87 \mathrm{nb}/ E (\mathrm{GeV})^2$
and $n = 8\times 10^{29}$ cm$^{-3}$.  The momenta randomisation 
  will become more efficient considering the radiation field
generated by the damping of the electrons in the magnetic field.  To
calculate the temperature of this electron-photon plasma we assume
that all the initial magnetic energy remains confined in the same
  volume of the vacuum break-down.  This density corresponds to
  a radiation gas with a temperature
$$
T_0 = \left( \frac {B^2}{8 \pi} \cdot \frac{1}{a} \right)^{1/4} \sim 10^{10} {\rm K}.
$$
The corresponding mean free path in the comoving frame, using
$\sigma_{\rm compt} \sim 1/3 \sigma_{\rm T} \sim 2 \times 10^{-25}
{\rm cm}^{2}$ and a radiation density $\sim 5 \times 10^{31}$
cm$^{-3}$ is around $l \sim 10^{-7}$ cm. The observed time scale
$t_{\mathrm{rand}}$ for this process is then
$$t_{\rm rand} \sim 10^{-16} \eta_{\rm acc}\qquad {\rm s}.$$

The energy of the particles in the plasmoid before the cooling by
synchrotron emission is
 
$$ E_{\mathrm{plasmoid}} = V \frac{B^2}{8 \pi} \sim 10^{45} 
\mathrm{erg}.$$

The available energy in the overall inelastic collision is $\Delta E
\sim 10^{53}$ erg, so that the emission of plasmoids could happen
$N_{\mathrm{plasmoid}}$ times where:
 
$$N_{\mathrm{plasmoid}} \sim \eta_{\rm B} \frac {\Delta
  E}{E_{\mathrm{plasmoid}}} = 10^8\, \eta_{\rm B},$$
where we have taken into
account also an efficiency, $\eta_{\rm B}$, for conversion of mechanical
energy into the electro-magnetically generated e$^{\pm}$ fireball.

The model therefore predicts for long duration GRB a pulsed emission
from $\sim$ 10$^7\, \eta_{\rm B,0.1}$ emitted plasmoids with an average
time separation $\Delta t \sim t_{\mathrm{obs}}/N_{\mathrm{plasmoids}}
\sim 10^{-5}$ s, corresponding to a separation distance $\sim 3 \times
10^5$ cm.  This large number of shells are likely to merge, thus
producing a significantly smaller number of well defined spikes in the
light curve with superposed a low amplitude flickering due to
individual shells.  The train of ``sausage'' plasmoids is $3 \times
10^{12} $cm long at the end of engine activity, even if its length
could be slightly modified during the internal shock phase.

The overall time duration is dominated by the duration of the engine
activity, the shortest variability time instead is determined by the
plasmoids interactions.  Variations in the observed luminosity
can be due e.g. to the very same formation of (internal) shocks among
the plasmoids and/or fluctuations in the accretion/field intensity.

\section{Collimation and acceleration: Two phase expansion}
 
As already discussed under the fermion pressure the fireball would
expand but the transversal motions are damped by the residual $B$
which provides the e$^{\pm}$ plasma confinement and collimation and is
responsible for synchrotron radiation. Thus the only motion possible
for the plasma bunch is that parallel to $B$: the plasmoid thus
becomes a stream of particles with velocity parallel to the external
field lines (see Section 2.3) with a corresponding bulk Lorentz factor
$\Gamma_1 \sim 30\, \eta_{\rm acc}$.
 
Within this scenario \footnote {We note that the collimation and
acceleration process discussed in the following does qualitatively
apply to a range of scenarios wider than that discussed in the
previous section.}, here we introduce the simplifying hypothesis that
the jet evolution (as recalled above) is composed by two distinct
phases, the first one (phase-1), occurring close to the engine
responsible of energising and collimating the burst.  Phase-1 ends (at
$R_1$) when the pre--existent collimating mechanism (e.g. in the case
of magnetic confinement the pre--existent magnetic field) cannot
balance the jet pressure any further.  We could give a rough
estimation of $R_1$ considering the distance when the collimation time
scale (for particles with $p_{\perp} \propto k T_0$) becomes equal to
the randomisation time scale. Following the discussion of the previous
Section, this happens when the external magnetic field is decayed to
$B\sim 10^{9}$ G. Assuming a dependence to $R^{-3}$ of the magnetic
field, $R_1$ could be estimated at a distance $\sim 10^8$ cm.

It then follows the second phase (phase-2), which consists of
adiabatic expansion and corresponding acceleration of the relativistic
particle fluid. This phase lasts for the radiation dominated phase and
ends at the smaller of the two radii (e.g. Piran 1999):
$$R_{\eta} = R_0 \frac{E}{Mc^2}$$
$$R_{\rm pair} = \left( \frac{3 E}{4\pi R_0^3 a T_p^4} \right)^{1/4}
R_0, $$
where R$_0$ is the initial source radius of the order of $R_s$, $R_{\eta}$ is the radius where the
fireball becomes matter dominated and $R_{\rm pair}$ where it becomes 
optically thin to pairs, corresponding to $T_p$ around 20 keV. 
Assuming that the total mass of the shell is dominated by the
electrons, which is justified by the very low environment density 
  (estimated in Section 2.2 as $\sim 10^{8}$ cm$^{-3}$), these radii could
  be estimated as
  $$R_{\eta} \sim 100 R_0 \qquad\qquad {\rm and} \qquad\qquad R_{\rm
    pair} \sim 50 R_0.$$

Therefore, for a radiation dominated expansion (Paczynski 1986):
$$
\frac{\Gamma_2^{'}}{\Gamma_1^{'}} \sim \frac{R_{\rm pair}}{R_0} \sim 50,
$$ 
where $\Gamma_1^{'} \sim 2$ is derived by the mean energy after the
collimation phase measured in the comoving frame of the collimated fireball 
shell, $\Gamma_2^{'}$ is the Lorentz factor at the end of
phase 2 measured in the same reference frame.

At the moment of transparency the ejecta are moving according to the
relativistic velocity composition in such a way that a particle
accelerated during the radiation dominated expansion in the
collimation direction will have
 
$$
\Gamma_{\parallel} = 2 \Gamma_1 \Gamma_2^{'} $$
where
$\Gamma_{\parallel}$ is the bulk Lorentz factor in the axis direction
assuming $\Gamma_1$ as the Lorentz factor of the moving shell in the
observer frame.
 
The opening angle of the conical jet structure generated at the end of
the two phases is determined by the particles accelerated
perpendicularly to the collimation moving with Lorentz factor
$\Gamma_{\perp} = \Gamma_2^{'}$. The angle $\theta_{\rm c}$ with
respect to the collimation axis is then:
 
$$ \theta_{\rm c} \sim \tan \theta_{\rm c} = \frac{\Gamma_{\perp} }{ \Gamma_1
\Gamma_2^{'} }= \frac{\Gamma_2^{'}}{ \Gamma_1 \Gamma_2^{'}} =
\frac{1}{\Gamma_1 }. $$

Assuming $\Gamma_{\parallel} \sim 10^3$ (e.g. Lithwick \& Sari 2001
for recent lower limit estimates) and estimating $\Gamma_2^{'}
\sim 100$ from the previous considerations, the value $\Gamma_1$ at
the end of the collimation phase has to be of the order of $ \Gamma_1
\sim 5 $ and consequently
$$
\theta_{\rm c} \sim \frac{1}{\Gamma_1} \sim 2 \times 10^{-1}.
$$

Values compatible with this estimate of $\Gamma_1$ have been derived
in Section 2.3, with reasonable assumptions on $\eta_{acc}\sim 0.1$.

From the arguments presented here it follows that if the ejected
shells are constituted by electron-positron pairs, travelling almost
parallel at the end of phase-1, their internal energy, as provided by
the central engine, corresponds to $\sim$ 1 MeV. The corresponding
collimation is within an angle of order of a few degrees.

\begin{figure}
\centerline{\psfig{figure=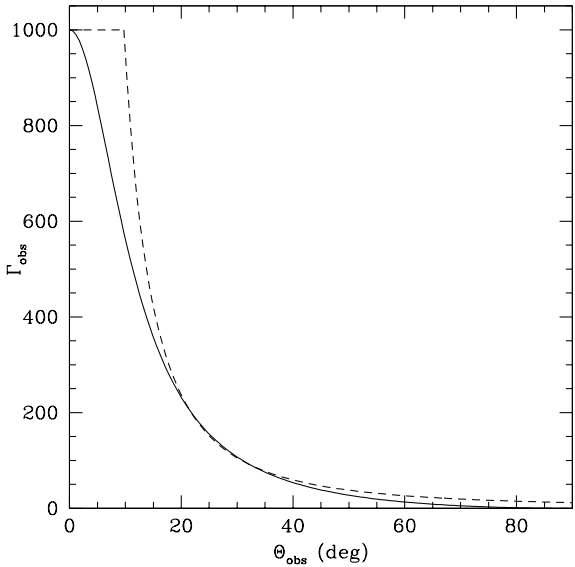,width=0.45\textwidth}}
\caption{ Predicted dependence of the energy as a function of the 
  observing angle $\theta$ for $\Gamma_1 = 5$, $\Gamma_2' = 100$ and
  $\theta_{\rm c} \sim \Gamma_1^{-1} $(solid line). For comparison we
  report as an example the indicative behaviour postulated by Rossi et
  al. (2002) for the corresponding value of $\theta_{\rm c} =
  10^{\circ}$.}
\end{figure}

Furthermore, in the above scenario, the observed angular distribution
of the expanding fireball is expected to be anisotropic and in
particular is simply given by the following expression:
  $$
  \tan(\theta) = \frac{\sin \theta^{'}}{\Gamma_1 (\beta_1 + \cos \theta^{'})},
  $$
  where $\theta^{'}$ is the angle of the emitted particle in the
  frame of the expanding fireball.

The observed energy of the same particle could then be estimated as:
$$
E(\theta') \propto \Gamma_1 \Gamma_2^{'} (1 + \beta_2' \cos
\theta^{'}).
$$
This angular distribution of the fireball Lorentz factor is shown
in Fig.~1.  Clearly, it is (qualitatively) expected a significant bias
in the selection of bursts observed with a specific instrument (i.e. a
defined energy window) not only because of flux but also peak energy
limits. Note also that, intriguingly, the dependence of $E(\theta')$
is qualitatively similar to that postulated by Zhang \& Meszaros
(2002), Rossi et al. (2002) (reported as a reference example in
Figure~1) to account for the phenomenological findings by Frail et al.
(2001).  It is also worth recalling the recent results by
Lloyd--Ronning \& Ramirez--Ruiz (2002) and Salmonson \& Galama (2002)
who find that observational (spectral and temporal) trends are better
accounted for in models where the burst anisotropy can be ascribed to
a dependence of the Lorentz factor (rather than barion loading) with
angle from the jet axis, in agreement with the most direct predictions
of the proposed scenario.

\section{Conclusions}

We considered the possibility that fireballs in long GRB are created
by a high magnetic field that breaks down the vacuum around the BH and
gives origin to a e$^\pm$ fireball. The energy can be extracted from a
rotating BH via the Blandford-Znajek mechanism thanks to a strong
magnetic field generated by an accreting torus.

The fireball evolution should then proceed in two phases, the first
one consisting in the energisation and collimation of the shells by
the external magnetic field and the second one - a radiation dominated
expansion - corresponding to the acceleration of the relativistic
photon--particle fluid and ending at the transparency radius.  This
scenario predicts that the resulting Lorentz factor is determined by
the product of the Lorentz factor of the shell close to the engine and
the Lorentz factor derived by the expansion and simply leads to the
formation of an anisotropic fireball.  For typical parameters expected
in the model the opening angle of the jet obtained in this model could
be then estimated to be of order of a few degrees, depending on the
efficiency of the acceleration and the resulting angular dependence is
similar to what already proposed in the literature on different
grounds.

\section{Acknowledgements}

We wish to thank Luigi Stella for providing the name of this model
during the first Italian national congress on GRB. AC also thanks
Enrico Ramirez--Ruiz for stimulating conversations. The Italian MIUR
is thanked for financial support (AC).

\end{document}